# Comment on "Astronomical alignments as the cause of ~M6+ seismicity"

Damián H. Zanette*

*The Patagonian Royal Society*

**Abstract.** It is shown that, according to the criteria used by M. Omerbashich (arXiv:1104.2036v4 [physics.gen-ph]), during 2010 the Earth was aligned with at least one pair of planets some 98.6% of the time. This firmly supports Omerbashich's claim that 2010 strongest earthquakes occurred during such astronomical alignments. On this basis, we argue that seismicity is, generally, a phenomenon of astrological origin.

*Also at Centro Atómico Bariloche and Instituto Balseiro, 8400 Bariloche, Río Negro, Argentina. Email: zanette@cab.cnea.gov.ar

In a very recent preprint [1], M. Omerbashich provides empirical evidence of a significant correlation between alignment events between our planet and other heavenly bodies of the Solar System, and strong earthquakes in the Earth. Specifically, he shows the complete coincidence between all recorded earthquakes of magnitude $M > 6$ during 2010, and such astronomical alignments. A similar comparison is provided for very strong earthquakes ($M > 8$) in the period 2000-2011, and for the 11 top strongest earthquakes ($M > 8.6$) recorded from 1902. In all cases, it is possible to identify at least two other components of the Solar System which were aligned with the Earth at the moment of the seismic event. The set of heavenly bodies considered in this study comprised the Sun, the Moon, the major planets (excluding the recently degraded "planetoid" Pluto) and, for earthquakes occurred since 2007, the comet C/2010 X1 (Elenin).

The statistical significance of these remarkable empirical observations, and their causal implications for earthly seismicity, can be enhanced by a more systematic, yet elementary analysis of the occurrence of astronomical alignments along a given period. In fact, while Omerbashich's work is exhaustive with respect to the record of earthquakes above a certain magnitude, it doesn't inform how frequently an alignment is effectively associated with an earthquake.

The alignment of our planet with other two astronomical bodies occurs when, in the course of their respective motions, the three bodies are approximately arranged along a straight line. During an alignment, as seen from the Earth, the two other objects appear either close to each other in the sky (when our planet is aside the two objects) or in practically opposite positions (when our planet lies between them). A quantitative assessment of the alignment's "quality" is given by the angular distance between the directions from the Earth to the two bodies. Specifically, if $\delta$ is the angle between the two directions ($0°<\delta<180°$), we can define the *alignment amplitude* as

$$\Delta = \min\{\delta, 180°-\delta\},$$

which takes into account the two possible relative positions of the Earth. The alignment effectively occurs if $\Delta$ is below a given *threshold* $\Delta_{max}$.

Unfortunately, Omerbashich does not specify which value of $\Delta_{max}$ was used in his study (or, as a matter of fact, how alignments were defined). Analyzing his data, however, it is possible to infer information about this crucial point. Figure 1 shows a histogram of the amplitude $\Delta$ for the 94 alignments related to strong 2010 earthquakes (Table 1 of Ref. 1) –excluding, as justified below, those which involve the Moon and the comet Elenin. To calculate these amplitudes, we used the heliocentric ecliptic coordinates of the planets (from Mercury to Neptune) for every day during year 2010, as provided in http://cohoweb.gsfc.nasa.gov/helios/planet.html. The coordinates for the Moon and the comet Elenin were not available and, therefore, these two bodies were not taken into account in our calculations.

The histogram shows that, while a substantial number of alignments have amplitudes below 5°, there is also a generous tail, almost reaching 20°. The largest amplitudes correspond to the January 2010 alignment of our planet with Mars and the Sun (with the

Earth between the two other objects), related to a "swarm" of thousands of seismic events recorder in Yellowstone, U. S. A. [1]. The average amplitude over the set of alignments considered in Fig. 1 is $\Delta_{av} = 4.95$ °.

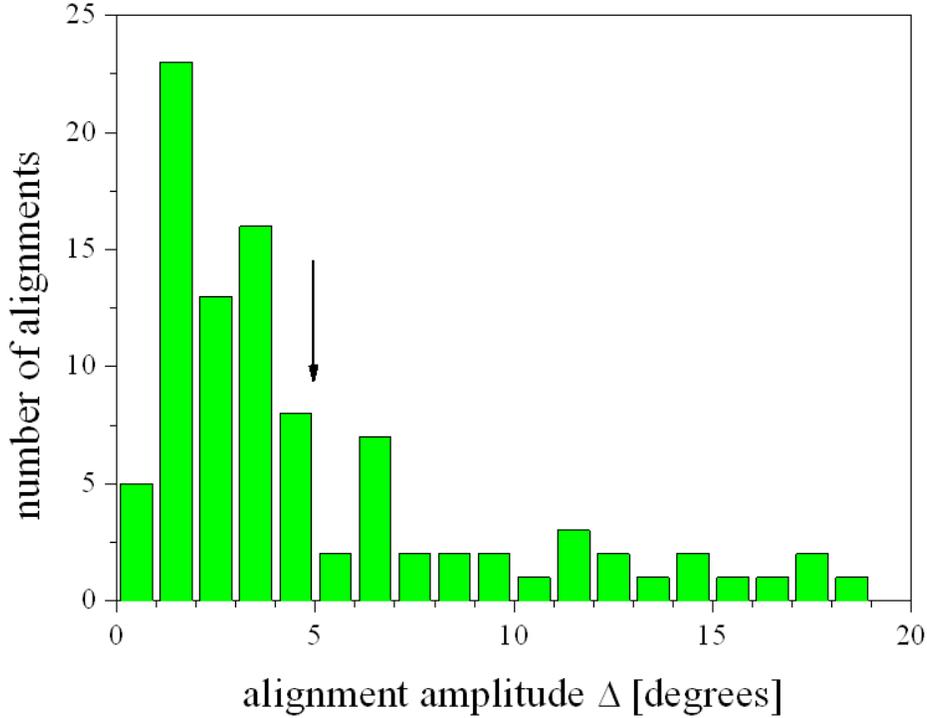

**Figure 1.** Histogram of the amplitudes for the 94 alignment events reported in Table 1 of Ref. 1 (excluded those involving the Moon and the comet Elenin). The arrow indicates the average amplitude, $\Delta_{av} = 4.95$ °.

The same heliocentric coordinates used above to calculate the amplitudes $\Delta$, make it possible to evaluate the frequency of the alignments. Specifically, we are interested at determining during how many days in year 2010 the Earth was aligned with at least two other objects. Naturally, this number depends on the threshold $\Delta_{max}$ used to define whether an alignment takes place or not. Figure 2 shows the fraction of the time in year 2010 –calculated as the number of days divided by 365– during which the Earth participated of at least one alignment, as a function of the threshold $\Delta_{max}$.

Taking as a statistically justifiable threshold the average amplitude $\Delta_{av} = 4.95$ ° derived from Omerbashich's data we find that, out of 365 days, the Earth was aligned with other bodies during 360 days or, in relative terms, 98.6% of the time. Only during the period from the 1[st] to the 4[th] of February and the 24[th] of the same month, was our planet alignment-free during the year 2010. In other words, under the present criterion, the probability that a strong 2010 earthquake, occurring at a randomly selected day, *did not* coincide with an astronomical alignment was as small as 0.014.

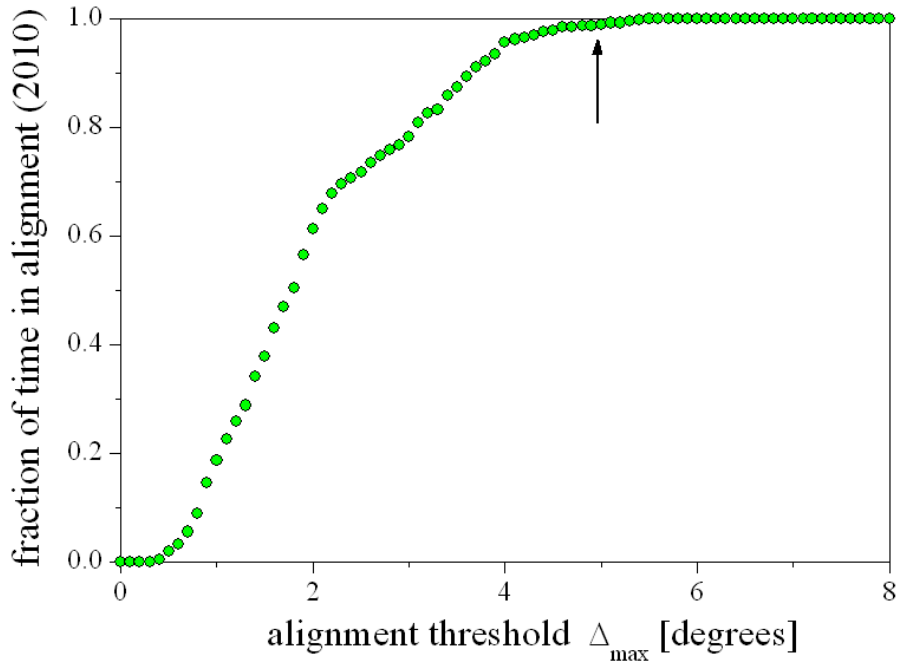

**Figure 2.** Fraction of time during which the Earth was participating of at least one alignment in year 2010, as a function of the threshold which defines alignment occurrences. The arrow indicates the average amplitude, $\Delta_{av} = 4.95$ °.

To our understanding, these results incontestably demonstrate that strong 2010 earthquakes were bound to occur during astronomical alignments. This, in turn, reinforces Omerbashich's hypothesis of a connection between such configurations and the seismic response of the Earth as a *georesonator* [1]. While the present results only apply to year 2010, it can be argued that qualitatively similar conclusions could be drawn for other periods, since 2010 had nothing special regarding the configuration of the Solar System. The inclusion of the Moon and other minor bodies such as the comet Elanin could only enhance the results' implications, as this would increase the number of alignments occurring in any given period –especially, because of the fast movement of these specific bodies across the sky.

As a final note, we would like to recall that, in 1687, Sir Isaac Newton proposed a rather convincing theory to explain the mutual influence of heavenly (and other) bodies under the action of the force of gravity [2] –the same force that, according to Omerbashich, underlies the effects of aligned objects on the Earth seismicity [1]. Newton's theory emphasizes the role of the mass and the distance between bodies in their interaction. Now, the only phenomena involving real-life objects that allegedly escape the effects of such factors are those of astrological nature. In these, geometrical configurations are far more relevant than mutual distances or masses. In the light of Omerbashich's results, complemented by our present contribution, it immediately follows that strong seismicity in our planet is –at least, to a large measure– a phenomenon of astrological origin. This conclusion seems to qualitatively agree with the traditional Homeric view that earthquakes were caused by the god of the seas, Poseidon (=Neptune, one of the planets considered in the above studies), shaking the Earth when he was angry.